\documentclass[unsortedaddress,superscriptaddress,prl]{revtex4-2}
\usepackage{hyperref}
\usepackage{epsfig}
\usepackage{graphicx}
\usepackage{subfigure}
\usepackage{latexsym}
\usepackage{color}
\usepackage{fullpage}
\usepackage{dcolumn}
\usepackage{bm}
\usepackage[normalem]{ulem}
\usepackage{units}
\usepackage{amsmath}
\usepackage[paperwidth=210mm,paperheight=297mm,centering,hmargin=2cm,vmargin=2.3cm]{geometry}

\begin{document}

\title{Synchrotron x-ray diffraction and DFT study of non-centrosymmetric EuRhGe$_3$ under high pressure}

\author{N. S. Dhami}
\email{nsdhami@ifs.hr}
\affiliation{Institute of Physics, Bijeni\v{c}ka cesta 46, 10000, Zagreb, Croatia}

\author{V. Balédent}
\affiliation{Université Paris-Saclay, CNRS, Laboratoire de Physique des Solides, 91405 Orsay, France}

\author{ I. Batistić}
\affiliation{Department of Physics, Faculty of Science, University of Zagreb, Bijeni\v{c}ka 32, 10000 Zagreb, Croatia}

\author{O. Bednarchuk}
\affiliation{Institute of Low Temperature and Structure Research, Polish Academy of Sciences, Okólna 2, 50-422 Wrocław, Poland}

\author{D. Kaczorowski}
\affiliation{Institute of Low Temperature and Structure Research, Polish Academy of Sciences, Okólna 2, 50-422 Wrocław, Poland}

\author{J. P. Itié}
\affiliation{Synchrotron SOLEIL, L’Orme des Merisiers,  91190 Saint-Aubin, France}

\author{S. R. Shieh}
\affiliation{Department of Earth Sciences, Department of Physics and Astronomy, University of Western Ontario, London, Ontario N6A-5B7, Canada}

\author{C. M. N. Kumar}
\affiliation{The Henryk Niewodniczański, Institute of Nuclear Physics, Polish Academy of Sciences, ul. Radzikowskiego 152, 31-342 Kraków, Poland}

\author{Y. Utsumi}
\email{yutsumi@ifs.hr}
\affiliation{Institute of Physics, Bijeni\v{c}ka cesta 46, 10000, Zagreb, Croatia}


\begin{abstract}
Antiferromagnetic intermetallic compound EuRhGe$_3$ crystalizes in a non-centrosymmetric BaNiSn$_3$-type ($I4mm$) structure. We studied its pressure-dependent crystal structure by using synchrotron powder x-ray diffraction at room temperature. Our results show a smooth contraction of the unit cell volume by applying pressure while preserving $I4mm$ symmetry. No structural transition was observed up to 35 GPa. By the equation of state fitting analysis, the bulk modulus and its pressure derivative were determined to be 73 (1) GPa and 5.5 (2), respectively. Furthermore, similar to the isostructural EuCoGe$_3$, an anisotropic compression of $a$ and $c$ lattice parameters was observed. Our experimental results show a good agreement with the pressure-dependent structural evolution expected from theoretical calculations below 13 GPa. Reflecting a strong deviation from integer Eu valence, the experimental volume data appear to be smaller than those of DFT calculated values at higher pressures.
\end{abstract}

\maketitle


\section{Introduction}

Rare-earth compounds in non-centrosymmetric BaNiSn$_3$-type structures have attracted considerable attention for their rich magnetic properties and pressure-induced superconductivity \cite{Smidman2017}. A series of Eu-based silicides/germanides Eu$TX_3$ (\textit{T}= transition metal, $X$=Si or Ge) have been discovered with the BaNiSn$_3$-type structure. In the crystallographic unit cell of Eu$TX_3$ systems, Eu atoms occupy the 2$a$ Wyckoff site, silicon/germanium atoms are located at two different Wyckoff positions 2$a$ and 4$b$, while transition metal atoms occupy the 2$a$ site \cite{Kaczorowski2012, Bednarchuk_JAC_2015}. Due to the lack of inversion symmetry in the crystal structure, the Dzyaloshinskii-Moriya antisymmetric spin interaction plays an active role and results in complex magnetic structures among the Eu$TX_3$ series \cite{Bednarchuk_JAC_2015,Bednarchuk_JAC_2_2015, Bednarchuk_APPA_2015, Maurya2014,Maurya2016, Matsumura2022}.

EuRhGe$_3$ possesses magnetic Eu$^{2+}$ (4f$^7$, $J$=7/2) ions and exhibits antiferromagnetic ordering in the tetragonal $ab$-plane below $T_{\rm N}$= 11.3 K \cite{Bednarchuk_JAC_2015,Bednarchuk_JAC_2_2015}. Temperature-dependent electrical resistivity of EuRhGe$_3$ shows a linear increase of $T_{\rm N}$ by applying pressure up to 8 GPa \cite{Kakihana2017}. Since the energy difference between Eu$^{2+}$ and nonmagnetic Eu$^{3+}$ (4f$^6$, $J$=0) is not very large \cite{Bauminger1973}, and the ionic radius of Eu$^{3+}$ is $\sim$ 10\% smaller than Eu$^{2+}$ ion, a pressure-induced Eu valence transition accompanied by a collapse of antiferromagnetic ordering and a volume change was expected. This phenomenon has been observed in ternary Eu-compounds with the ThCr$_2$Si$_2$-type ($I4/mmm$) structure \cite{Onuki_PM_2017, Onuki2020}. However, the antiferromagnetic ordering and divalent Eu were revealed to be stable against pressure in EuRhGe$_3$. Recently, we performed high energy resolution fluorescence detected (HERFD) near-edge x-ray absorption spectroscopy (XAS) on EuRhGe$_3$ as a function of pressure. In the Eu $L_3$-edge XAS spectra, a prominent Eu$^{2+}$ peak was observed at ambient pressure. By increasing pressure, the spectral intensity shifted from Eu$^{2+}$ to Eu$^{3+}$ peaks. The obtained mean Eu valence from the Eu $L_3$ XAS spectrum exhibited a continuous increase from $\sim$ 2.1 at ambient pressure to $\sim$ 2.4 at 40 GPa without a first-order valence transition \cite{Utsumi_ES_2021}. 

Pressure-dependent Eu valence and crystal structure were also studied in isostructural antiferromagnets EuCoGe$_3$ and EuNiGe$_3$. Their antiferromagnetic phases are stable against pressure \cite{Uchima2014, Kakihana2017, Muthu2019}. Both exhibited a continuous contraction of their unit cell volume by applying pressure without any symmetry changes \cite{dhami2023pressure, chen2023evidence}. The mean Eu valence in EuCoGe$_3$ only changes from 2.2 at 2 GPa to $\sim$ 2.3 even around 50 GPa \cite{dhami2023pressure}. In EuNiGe$_3$, the mean Eu valence changes from 2.13 at 1 GPa to 2.43 at 48 GPa at 8 K \cite{chen2023evidence}. As well as EuRhGe$_3$, the Eu valence does not reach Eu$^{3+}$ in both EuNiGe$_3$ and EuCoGe$_3$, even under such high pressure. These pressure-dependent changes of Eu valence in the Eu$T$Ge$_3$ series are rather small compared to ternary Eu-compounds with the ThCr$_2$Si$_2$-type structure (Eu122-systems), which often show a drastic pressure-induced Eu valence transition below 10 GPa \cite{Onuki_PM_2017, Onuki2020}. For example, antiferromagnet EuRh$_2$Si$_2$ exhibits a pressure-induced valence transition at room temperature. By applying pressure, the mean Eu valence changes from Eu$^{2.2+}$ at ambient pressure to almost Eu$^{3+}$ at $\sim$8 GPa \cite{mitsuda2018pressure}. Such pressure-induced valence transition is also reported in antiferromagnet EuNi$_2$Ge$_2$, where the Eu valence changes from 2.2 at ambient pressure to almost Eu$^{3+}$ above 5 GPa at room temperature \cite{Hess1997}. In the Eu122-systems, a correlation between the Eu valence and its unit cell volume tends to appear as a lattice volume collapse simultaneously happening with the Eu valence transition since the size of the Eu$^{3+}$ ion is about 10\% smaller than the Eu$^{2+}$ ion \cite{Shannon1976}. 


The different pressure behavior in the Eu$T$Ge$_3$ series from the Eu122-systems raises questions about whether the variation of the Eu valence in the Eu$T$Ge$_3$ series correlates with the pressure evolution as well as their crystal structural changes. Unlike the intensively investigated Eu122-systems, pressure-dependent crystal structural studies of the Eu$T$Ge$_3$ series remain scarce. To elucidate a trend of pressure-dependent structural changes in the Eu$T$Ge$_3$ series and its relation to the Eu valence evolutions, a systematic study is needed. Here we performed synchrotron powder x-ray diffraction (XRD) as a function of pressure to study the pressure-dependent crystal structural change in EuRhGe$_3$, together with the theoretical predictions by density functional theory (DFT) method. The results are discussed in relation to the pressure evolution of the mean Eu valence obtained by HERFD near-edge XAS \cite{Utsumi_ES_2021}.

\section{Experiment}
Single crystals of EuRhGe$_3$ were synthesized by the metal-flux method using liquid indium as solvent. The crystals were taken from the same batch reported in Ref.\cite{Bednarchuk_JAC_2015}. The grown crystals were characterized by XRD, electrical resistivity, and magnetic susceptibility measurement \cite{Bednarchuk_JAC_2015,  Bednarchuk_JAC_2_2015, Bednarchuk_APPA_2015}.

The high-resolution XRD was performed at the PSICHE beamline of SOLEIL synchrotron with a photon energy of 33 keV ($\lambda$ = 0.3738 \AA). The single crystal of EuRhGe$_3$ was ground into a powder using mortar and pestle with ethanol. Fine grains of EuRhGe$_3$ afloat on the ethanol were collected by using a syringe and transferred onto a clean glass slide. After drying the powdered sample was loaded in a membrane diamond anvil cell (DAC). Diamonds of 300 $\mu$m diameter culet size were used. A rhenium gasket was pre-indented to about 28 $\mu$m thickness, and then a hole of 150 $\mu$m was drilled through it to serve as a sample chamber. Gold powder was also loaded in the DAC as a pressure reference material. Helium was used as a pressure-transmitting medium. Pressure was controlled by a membrane on the DAC and was determined by the Au equation of state \cite{heinz1984}.

The pressure evolution of the crystal structure of EuRhGe$_3$
was also studied theoretically by using the Quantum ESPRESSO DFT package \cite{Giannozzi2009,Giannozzi2017}. In DFT calculations, we have used pseudopotentials from pslibrary 1.0.0 \cite{DALCORSO2014337}, with the Perdew–Burke–Ernzerhof exchange-correlation functional appropriate for solids \cite{Perdew2008, Perdew2009}. The kinetic energy cutoff for wavefunctions was 150 Ry, while for the charge density and potential, it was 700 Ry.
The Brillouin sampling was 16x16x8 (no offset), with the Marzari–Vanderbilt Fermi surface smearing \cite{Marzari1999}. To take into account the antiferromagnetic ordering on Eu,
we have used the simplified formulation of DFT+$U$ proposed by Dudarev \cite{Dudarev1998}.
The Hubbard interaction $U$ for Eu was assumed to be 3.8 eV to match the Eu 4$f$ peak in the valence band spectrum by photoelectron spectroscopy measurement \cite{Utsumi2018}. This $U$ value was kept constant for all pressure calculations.

\section{Results and Discussion}

\begin{figure*}[!t]
    \centering
    \includegraphics[width=\textwidth]{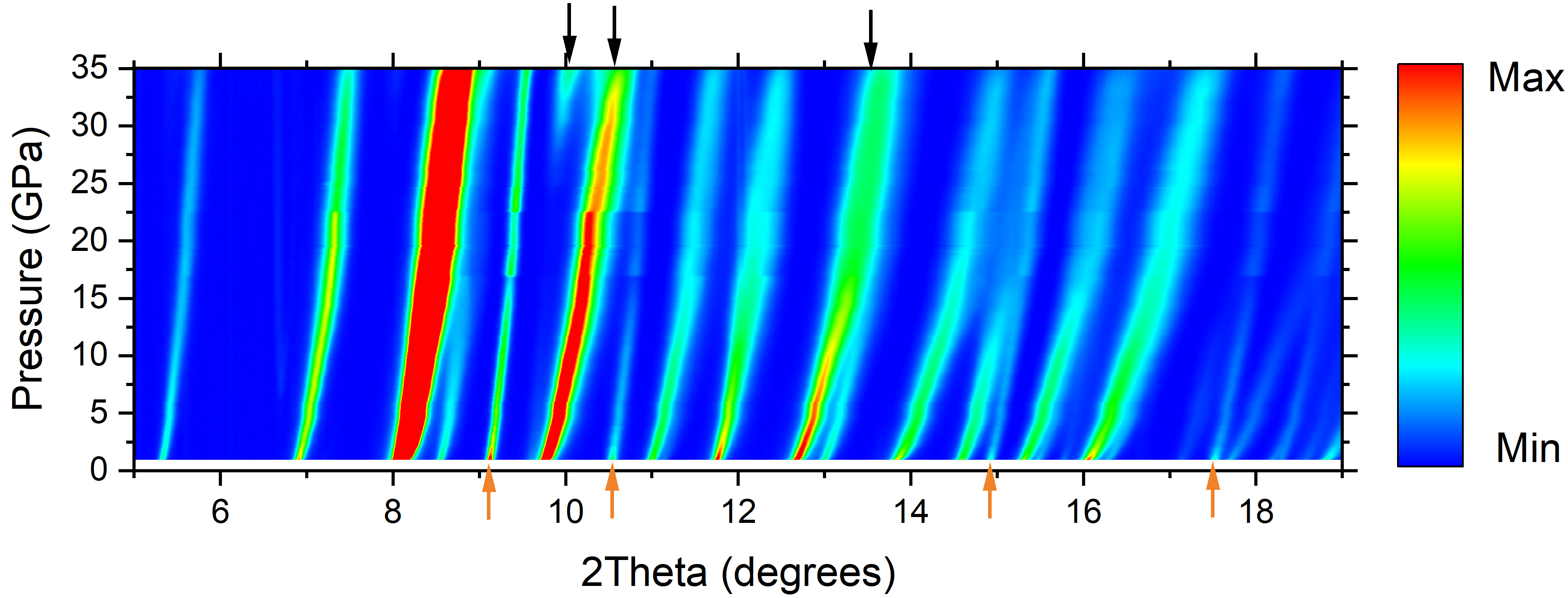}
    \caption{Contour map of synchrotron x-ray diffraction intensities in the pressure range {1--35~GPa}. The pressure evolution of Bragg peak positions contains EuRhGe$_3$ (main phase), gold (standard material), and rhenium (gasket) which appears above 25 GPa. The black arrows (top) and yellow arrows (bottom) indicate the main peak positions of rhenium and gold respectively.}
    \label{fig1}
\end{figure*}

We performed powder XRD on EuRhGe$_3$ as a function of pressure at room temperature up to 35 GPa. The contour map of diffraction intensities in the pressure range from 1 to 35 GPa is presented in Fig.{\ref{fig1}}. In order to highlight the pressure evolution of the peak positions, the contour map is plotted in the 2$\theta$ range from 5 to 19 $\deg$. The pressure evolution of Bragg peak positions contains the main phase EuRhGe$_3$ and gold as pressure reference. When pressure exceeds 25 GPa, new peaks that belong to neither EuRhGe$_3$ nor gold emerge in the diffraction pattern. These new peaks were identified as rhenium from the gasket. 

\begin{figure*}[!htb]
    \centering
    \includegraphics[width=0.6\textwidth]{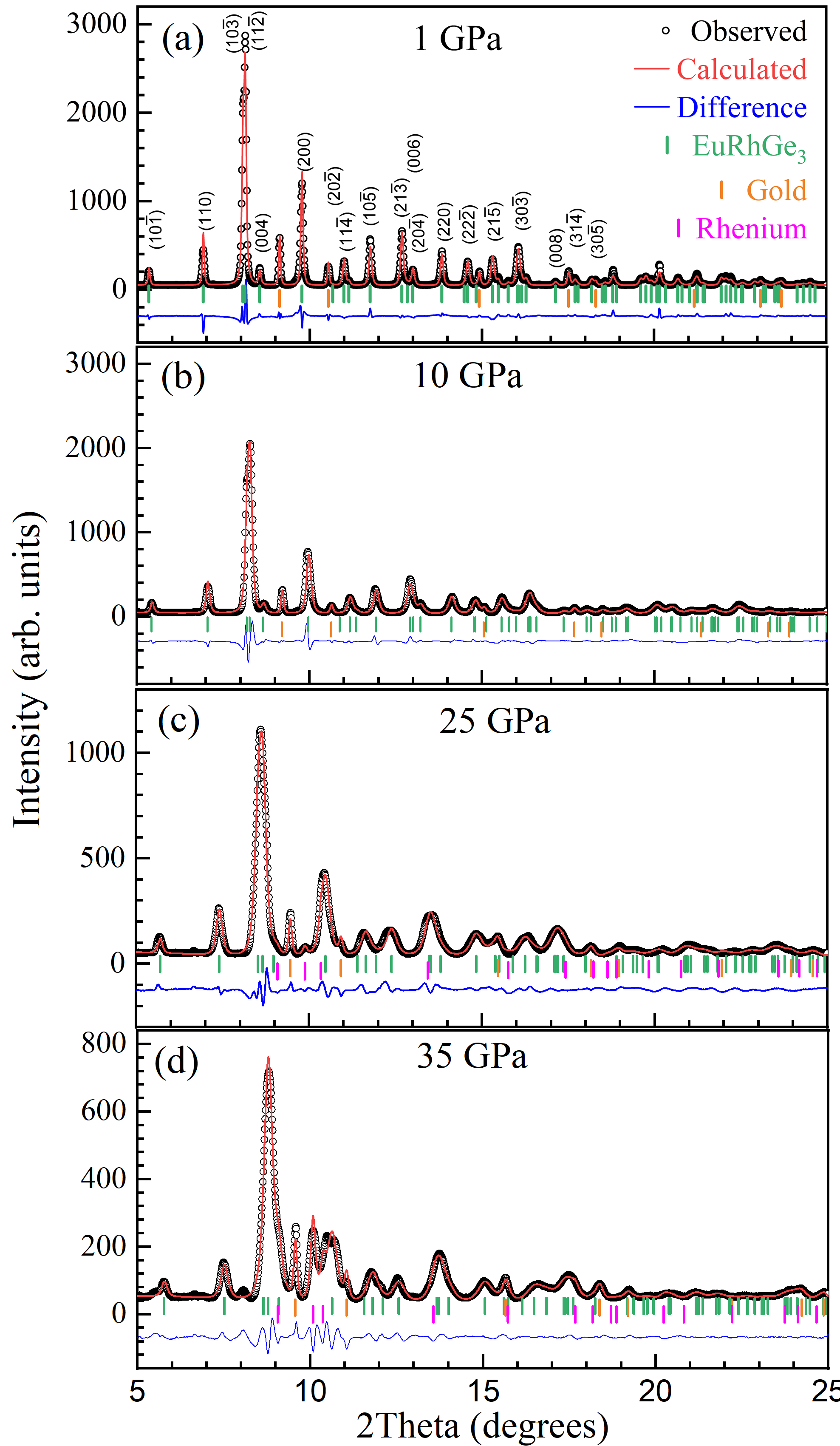}
    \caption{Synchrotron powder XRD patterns of EuRhGe$_3$ along with the results of Rietveld refinement at (a) 1 GPa, (b) 10 GPa, (c) 25 GPa, and (d) 35 GPa. The vertical bars indicate Bragg peak positions of EuRhGe$_3$ (green), gold (orange), and rhenium (magenta).}
    \label{fig2}
\end{figure*}

We integrated the diffraction images by using Dioptas software \cite{Dioptas} and performed Rietveld refinement for three phases by using the Profex program \cite{Profex}. The Rietveld refinement fits at selected pressures are presented in Fig. {\ref{fig2}}. The widening of diffraction peaks with increasing pressure might reflect the presence of non-hydrostaticity. Another plausible reason for the peak widening is the bridging of the powder sample between the anvils. Although the DAC was prepared not to have an excessive amount of the sample in the gasket hole, a possible occurrence of bridging cannot be eliminated. Note that, helium pressure medium has the highest hydrostatic limit among other rare gases \cite{takemura2001}, a development of uniaxial stress is reported above 30 GPa that largely depends on experiments \cite{takemura2007,Takemura2008}. The obtained lattice parameters and refinement parameters at selected pressures are shown in Table \ref {table1}. The detailed atomic coordinates as a function of pressure are presented in Appendix B. Our analysis confirms that EuRhGe$_3$ maintains the same crystal symmetry ($I4mm$) up to 35 GPa. Note that the unit cell parameters \textit{a} = 4.4129 Å, \textit{c} = 10.0906 Å,  and unit cell volume $V_0$ = 196.50 Å$^3$ at ambient pressure are taken from Ref. \cite{Bednarchuk_JAC_2015}.

\begin{table*}[htb!]
\centering
\begin{tabular}{c c c c c}
 \hline
 Pressure & 1 GPa & 10 GPa & 25 GPa & 35 GPa \\ [1.2ex]
 \hline
(Experimental)\\
 Lattice parameters and volume\\

a  (Å) & 4.3899 (1) & 4.249 (4) & 4.103 (2) & 4.04 (1)\\
c  (Å) & 10.035 (3) & 9.821 (9) & 9.567 (6) & 9.45 (2)\\
 V (Å$^3$) & 193.38 (9)  & 177.3 (3) & 161.1 (1) & 154.2 (8) \\
Refinement parameters\\
R$_{WP}$ & 9.47 & 10.39 &9.8 & 9.21\\
R$_{exp}$ & 10.11 & 9.55 & 9.98 & 10.52\\
$\chi^2$& 0.88 & 1.17 & 0.96 & 0.77\\
GOF& 0.94 & 1.06 & 0.98 & 0.91\\  
\hline
 Pressure & 0 GPa & 10 GPa & 20 GPa & 30 GPa \\ [1.2ex]
 \hline
(DFT)\\
 Lattice parameters and volume \\
a  (Å) & 4.3889 & 4.2547 & 4.1564 & 4.0774 \\
c  (Å) & 10.0478 & 9.8246 & 9.6787 &  9.5731 \\
 V (Å$^3$) & 193.5452 & 177.8495 & 167.2059 & 159.1546\\
\hline
\end{tabular}
\caption{Experimental lattice parameters, unit cell volume, and refinement parameters of EuRhGe$_3$ at selected pressures. Where R$_{WP}$ is the weighted profile R factor, R$_{exp}$ is the expected R factor and GOF is the goodness of fitting. Lattice parameters and unit cell volume obtained by DFT calculation at various pressures.}
\label{table1}
\end{table*}

\begin{figure*}[!h]
    \centering
    \includegraphics[width=\textwidth]{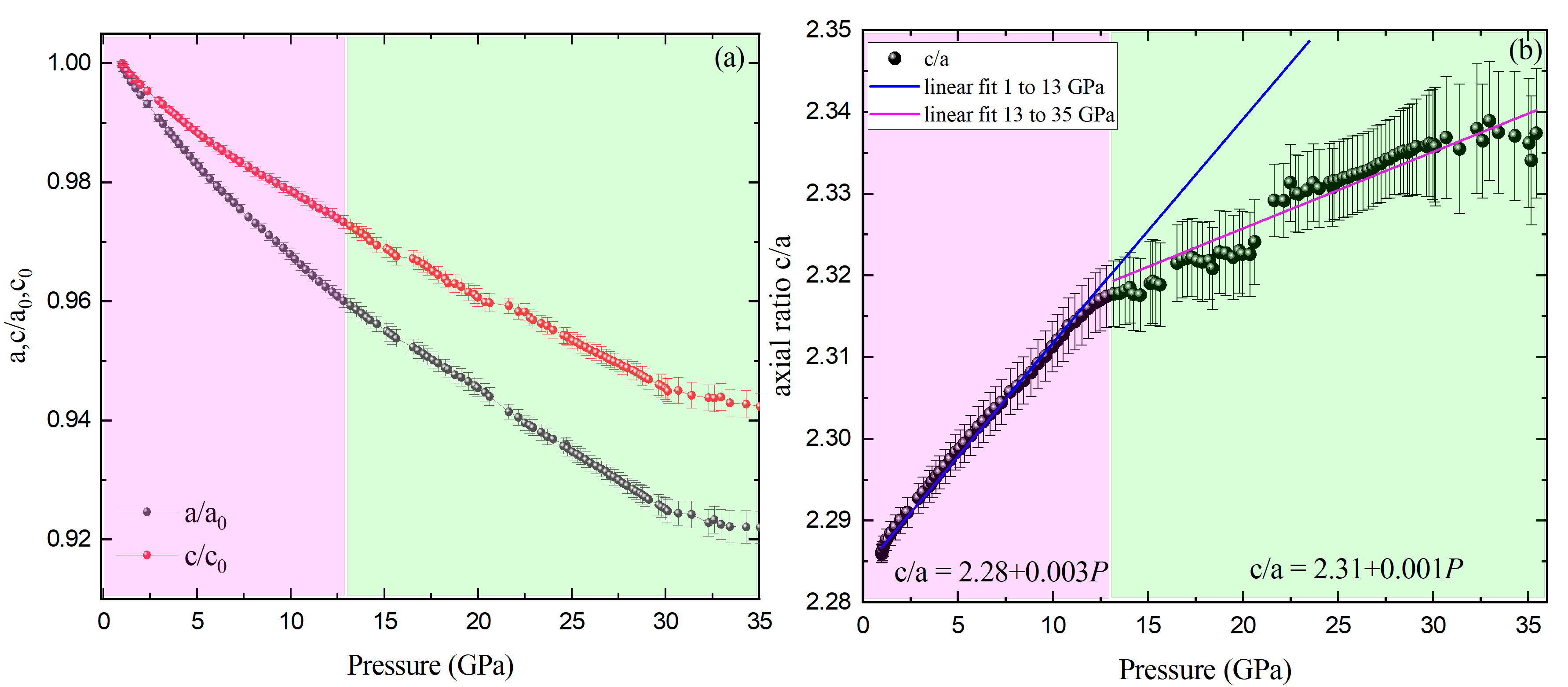}
    \caption{(a) Relative pressure variations of the lattice parameters of EuRhGe$_3$ with respect to the values at the lowest experimental pressure. (b) Pressure dependence of the axial ratio ($c/a$). Straight lines emphasize linear behavior. The two different colors represent two different pressure regions. To emphasize the change of slope, the linear fit of the LP region is extended into the HP region.}
    \label{fig3}
\end{figure*}

Figure \ref{fig3} (a) shows the evolution of lattice parameters \textit{a}- and \textit{c} as a function of pressure. The lattice parameter $a$ shows greater pressure change compared to $c$. Similar high-pressure behavior is also observed in EuCoGe$_3$ \cite{dhami2023pressure} and EuNiGe$_3$ \cite{chen2023evidence}. We also performed high-pressure powder XRD of EuRhGe$_3$ by using a neon pressure transmitting medium . The pressure-dependent changes of $a$ and $c$ lattice parameters showed the same tendency (see supporting material), which suggests an intrinsic pressure behavior to the Eu$T$Ge$_3$ series and is independent from the pressure medium. Figure \ref{fig3} (b) shows the change of axial ratio ($c/a$) as a function of pressure. Pressure evolution of the $c/a$ can be divided into two regions: low-pressure (LP) region \textless13 GPa and high-pressure (HP) region above 13 GPa. The linear fitting of the LP region provides 2.28+0.003\textit{P}, while it provides 2.31+0.001\textit{P} in the HP region. In the LP region, the value of the $c/a$ linearly increases with pressure with a rate of 0.003 GPa$^{-1}$, then the rate decreases to 0.001 GPa$^{-1}$ in the HP region. This change in the $c/a$ increase rate is ascribed to a change in compressibility along the $a$- and $c$-axis. The linear modulus of the $c$-axis increases considerably in the HP region, which makes the $c$-axis less compressible and changes the slope of $c/a$ (see Appendix A). Among Eu122-systems, the change of $c/a$ slope as a function of pressure was reported as a consequence of pressure-induced isostructural transition. Across such isostructural transition from tetragonal to so-called collapsed tetragonal phase, $a$-, $c$-lattice parameters and the unit cell volume greatly change along with the $c/a$ slope \cite{Hess1997, huhnt1997first, Huhnt1998, bishop2010, uhoya2010}. Compared to the large structural changes in Eu122-systems, no discontinuous changes are observed in lattice parameters at the pressure range of the $c/a$ slope change in EuRhGe$_3$. Moreover, the $c/a$ slope change does not coincide with the pressure change of Eu valence in EuRhGe$_3$. The isostructural phase transitions in the Eu122-systems are accompanied by pressure-induced Eu valence transition from almost Eu$^{2+}$ to Eu$^{3+}$. However, in EuRhGe$_3$, the mean Eu valence linearly increases from $\sim$ 2.1 at ambient pressure to $\sim$ 2.4 around 25 GPa, then the increase rate becomes smaller and deviates from the linear behavior above 25 GPa \cite{Utsumi_ES_2021}. Hence, we eliminate the possibility of pressure-induced isostructural transition in EuRhGe$_3$. It is worth mentioning that the change of the $c/a$ slope as a function of pressure was also observed in EuCoGe$_3$ \cite{dhami2023pressure} and EuNiGe$_3$ \cite{chen2023evidence} without any isostructural transitions.

\begin{table*}[htb!]
\centering
\begin{tabular}{c c c c}
 \hline
Compound & Unit cell volume (V$_0$) & Bulk modulus (B$_{0}$) & B$_{0}^\prime$\\

 \hline
EuCoGe$_3$ & 184.39 (Å$^3$)& 75.6 & 5.58 \cite{dhami2023pressure}\\ 
EuNiGe$_3$  & 185.7 (Å$^3$)& 79 & 8.8 \cite{chen2023evidence}\\ 
EuRhGe$_3$ (XRD) & 196.50 (Å$^3$)& 73 (1) & 5.5(2)\\
EuRhGe$_3$ (DFT) & 193.545 (Å$^3$)& 97.4(1) &4.61(1) \\
\hline
\end{tabular}
\caption{The unit cell volumes, bulk modulus, and first pressure derivative of bulk modulus of Eu$T$Ge$_3$ series and those obtained by the DFT calculation in EuRhGe$_3$. The values of EuCoGe$_3$ and EuNiGe$_3$ are taken from Ref. \cite{dhami2023pressure} and \cite{chen2023evidence}, respectively. }
\label{table2}
\end{table*}

\begin{figure*}[!htb]
    \centering
    \includegraphics[width=0.65\textwidth]{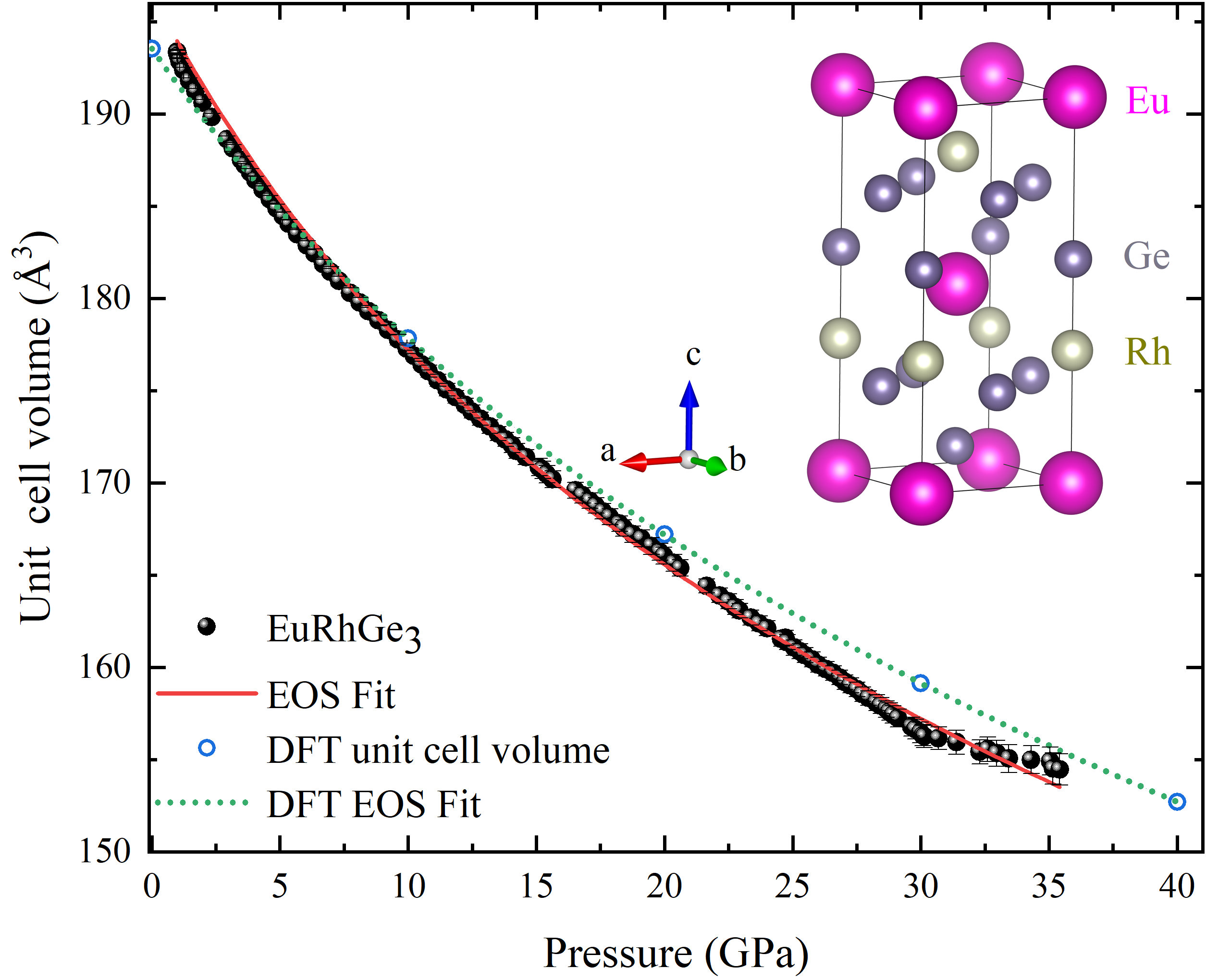}
    \caption{Pressure evolution of the unit cell volume of EuRhGe$_3$ (solid circles) with the result of EOS fitting (solid line). The unit cell volume obtained by DFT calculation for the selected pressures (open circles) and its EOS fitting (dashed line). The error bar is the range of symbol size. The inset shows the crystal structure of EuRhGe$_3$ drawn by VESTA \cite{Momma2011}.}
    \label{fig4}
\end{figure*}

Figure \ref{fig4} shows the pressure evolution of the unit cell volume of EuRhGe$_3$. It exhibits a smooth contraction with increasing  pressure. A slight change in the slope can be seen around 30 GPa (also in the lattice parameters of Fig.3 (a)). This slope change of the unit cell volume coincides with the slope change of the mean Eu valence as a function of pressure. The mean Eu valence of EuRhGe$_3$ obtained by HERFD XAS exhibited a linear increase by applying pressure up to 25 GPa and exhibited a smaller change at higher pressure \cite{Utsumi_ES_2021}. Pressure-dependent unit cell volumes obtained by the DFT calculation are also plotted in Fig. 4. The theoretically calculated unit cell volumes exhibit a good agreement with experimental results in the LP region. However, the deviation from the experimental results increases in the HP region. The DFT-calculated unit cell volumes in the HP region tend to be larger than the experimental values. This is due to the change of the $c$-axis compressibility in the HP region, which is not captured by the DFT calculation. Since the DFT calculations cannot take into account the valence fluctuation, the Eu valence was considered to be Eu$^{2+}$ in the whole pressure range. In the LP range, the average Eu valence in EuRhGe$_3$ stayed close to Eu$^{2+}$ (2.1-2.2) \cite{Utsumi_ES_2021} resulting in a reasonable agreement between the experiment and the DFT calculation. However, the mean Eu valence increases toward 2.4 by further increasing pressure. Considering the different ionic radius between Eu$^{2+}$ and Eu$^{3+}$ (~10\% smaller), it is somewhat understandable that the experimental unit cell volume appears to be smaller than the DFT-calculated values. In order to study the elastic properties of EuRhGe$_3$, we performed equation of state (EOS) fitting by using EOSFit7c software \cite{EosFit}. We used the 3$^{rd}$ order Birch Murnaghan equation below for EOS fitting \cite{Birch1947} :

\[P(V)=  \frac{3B_0}{2} \left [\left (\frac{V_0}{V}\right )^{7/3} - \left (\frac{V_0}{V}\right )^{5/3} \right ] \left \{1+\frac{3}{4}(B_0'-4)\left [\left (\frac{V_0}{V}\right )^{2/3} -1\right ]\ \right \} \]

Here, $B_0$ and B$_{0}^\prime$ denote the bulk modulus at 0 GPa and its first pressure derivative, respectively. We also performed EOS fitting on the lattice volume of EuRhGe$_3$ obtained by the DFT calculation. The obtained values from the EOS fitting are listed in the table \ref{table2} and compared with the reported values of EuCoGe$_3$ and EuNiGe$_3$, as well as those obtained in the theoretical unit cell volume of EuRhGe$_3$.

\section{Conclusion}
We performed pressure-dependent synchrotron x-ray diffraction measurements on polycrystalline non-centrosymmetric EuRhGe$_3$ at room temperature. It revealed a smooth contraction of the lattice constants without any structural transition within the investigated pressure range. An anisotropic compressibility between the $a$- and $c$-axis was observed. The lattice parameter along the $a$-axis exhibits more sensitivity to pressure and greater contraction in comparison to that of the $c$-axis. The change of slope in the pressure-dependent $c/a$ ratio was observed around 13 GPa, which does not seem to have a direct correspondence to the pressure evolution of the mean Eu valence in EuRhGe$_3$. By performing the EOS fitting on the obtained unit cell volume, we determined the bulk modulus and its pressure derivative of EuRhGe$_3$. Our experimental results show a good agreement with the pressure-dependent structural changes expected from the DFT calculations at P\textless 13 GPa, and slightly deviates in the HP region. The experimental unit cell volume tends to be smaller than the theoretical values in the HP region as a consequence of the increased mean Eu valence in the HP region.

\section*{Acknowledgments}
The synchrotron powder XRD experiments under pressure were performed at the PSICHE beamline station (proposal Nos. 20191645 and 20220361). This work has been supported by the Croatian Science Foundation under project No. UIP-2019-04-2154, and in part under project No. IP-2020-02-9666. N.S.D. acknowledges financing from the Croatian Science Foundation under the “Young Researchers’ Career Development Project:” Project No. DOK-2018-09-9906. N.S.D. and Y.U. acknowledge support of the project from the Cryogenic Centre at the Institute of Physics—KaCIF (Grant No. KK.01.1.1.02.0012), co-financed by the Croatian Government and the European Union through the European Regional Development Fund—Competitiveness and Cohesion Operational Programme. Y. U. thanks Kai Chen for the helpful discussion and for sharing the experimental data of EuNiGe$_3$ high-pressure XRD.

\clearpage
\pagebreak

\bibliographystyle{apsrev}
                \bibliography{Biblography_EuRhGe3_XRD}

\global\long\def\appendixname{APPENDIX}
	\section{APPENDIX}

\appendix

\section{Linear EOS fitting}
As mentioned in the main text, the pressure evolution of the $c/a$ ratio can be divided into the low-pressure (LP) region $<$ 13 GPa and the high-pressure (HP) region above 13 GPa. We also performed a linear EOS fitting by using the EOSFit7c software \cite{EosFit}. The linear modulus is defined as: 
\[M_i = -x_i \left (\frac{\partial P}{\partial x}\right)_T\] Thus the relation between the linear modulus and the bulk modulus can be described as follows: \[B = \left( \frac{1}{M_1} +\frac{1}{M_2}+\frac{1}{M_3} \right)^{-1} \] 

Here, M$_1$, M$_2$, and M$_3$ are linear moduli along the \textit{a}-axis, \textit{b}-axis, and \textit{c}-axis respectively. For tetragonal symmetry M$_1$ and M$_2$ are the same. The lattice parameters along the $a$- and $c$-axis at 0 GPa (\textit{a} = 4.4129 Å, \textit{c} = 10.0906 Å) used for the linear EOS fitting were taken from Ref.\cite{Bednarchuk_JAC_2015}. The linear EOS fitting was performed separately in the LP and the HP regions. The obtained linear moduli along the \textit{a}- and the \textit{c}-axis both increase in the HP region. The linear modulus in the HP region especially increases  along the \textit{c}-axis (See Table III).
The bulk modulus calculated by using the aforementioned formula is consistent with the bulk modulus extracted from the pressure-dependent unit cell volume in the main text. 
\begin{table*}[htb!]
\centering
\begin{tabular}{c c c}
 \hline
axis (Pressure region) & Linear moduli (M$_{0}$) & M$_{0}^\prime$\\

 \hline
$a$ (LP)& 179 (1) & 19.9 (5)\\ 
$a$ (HP) &  205 (4) & 14.2 (6)\\ 
$c$ (LP) &  182 (4)& 61 (3)\\
$c$ (HP) &  303 (5) &16.4 (7)\\
\hline
\end{tabular}
\caption{The linear modulus at 0 GPa (M$_0$), and first pressure derivative of linear modulus (M$_0^\prime$) of EuRhGe$_3$.}
\label{table:2}
\end{table*}

\section{DFT-calculated and experimental atomic coordinates}
Table IV presents the DFT-calculated atomic coordinates, and table V shows the atomic coordinates extracted from the Rietveld refinement of the XRD data. 
\begin{table*}[htb!]
\centering
\begin{tabular}{c c c c c c }
 \hline
Atoms  & Symmetry & x & y & z  \\ [1.2ex]
 \hline
0 GPa \\

Eu  & 2a & 0.0& 0.0 & 0.0  \\
Rh& 2a & 0.0 & 0.0 &  	0.3488 \\
Ge1 & 4b & 0.0  &  0.5	 &  0.2435 \\
Ge2 & 2a & 0.0	 & 0.0	&  0.5864  \\

10 GPa \\

Eu  & 2a & 0.0& 0.0 & 0.0 \\
Rh& 2a & 0.0 & 0.0 &  	0.3514   \\
Ge1 & 4b & 0.0 & 0.5& 0.2423  \\
Ge2 & 2a & 0.0	 & 0.0	&  0.5886\\

20 GPa \\

Eu  & 2a & 0.0& 0.0 & 0.0   \\
Rh& 2a & 0.0 & 0.0 &  	0.3532  \\
Ge1 & 4b & 0.0 & 0.5& 0.2412  \\
Ge2 & 2a & 0.0	 & 0.0	&  0.5897 \\

30 GPa \\

Eu  & 2a & 0.0& 0.0 & 0.0   \\
Rh& 2a & 0.0 & 0.0 &  	0.3547  \\
Ge1 & 4b & 0.0 & 0.5& 0.2404 \\
Ge2 & 2a & 0.0	 & 0.0	&  0.5903\\

40 GPa \\

Eu  & 2a & 0.0& 0.0 & 0.0  \\
Rh& 2a & 0.0 & 0.0 &  	0.3557 \\
Ge1 & 4b & 0.0 & 0.5& 0.2394  \\
Ge2 & 2a & 0.0	 & 0.&  0.5905 \\

\hline

\end{tabular}
\caption{Atomic coordinates of EuRhGe$_3$ obtained by DFT calculation.}
\label{table:3}
\end{table*}

\begin{table*}[htb!]
\centering
\begin{tabular}{c c c c c }
 \hline
Atoms  & Symmetry & x & y & z  \\ [1.2ex]
 \hline
1 GPa \\

Eu  & 2a & 0.0& 0.0 & 0.0  \\
Rh& 2a & 0.0 & 0.0 &  0.35377$\pm$ 0.00044 \\
Ge1 & 4b & 0.0  &  0.5 &  0.24164$\pm$0.00065 \\
Ge2 & 2a & 0.0& 0.0&  0.58194$\pm$0.00078  \\

10 GPa \\

Eu  & 2a & 0.0& 0.0 & 0.0 \\
Rh& 2a & 0.0 & 0.0 & 0.35424$\pm$0.00055   \\
Ge1 & 4b & 0.0 & 0.5& 0.24045 $\pm$0.00085   \\
Ge2 & 2a & 0.0& 0.0&  0.58161 $\pm$0.00055\\

25 GPa \\

Eu  & 2a & 0.0& 0.0 & 0.0   \\
Rh& 2a & 0.0 & 0.0 & 0.35556$\pm$0.00085  \\
Ge1 & 4b & 0.0 & 0.5& 0.2440$\pm$0.0011  \\
Ge2 & 2a & 0.0 & 0.0&  0.5916$\pm$0.0013 \\

35  GPa \\

Eu  & 2a & 0.0& 0.0 & 0.0   \\
Rh& 2a & 0.0 & 0.0 &  0.3568$\pm$0.0021   \\
Ge1 & 4b & 0.0 & 0.5& 0.2462 $\pm$0.0011  \\
Ge2 & 2a & 0.0& 0.0&  0.593$\pm$0.0019\\

\hline

\end{tabular}
\caption{Experimental atomic coordinates of EuRhGe$_3$ extracted from the refinement of the XRD data.}
\label{table:4}
\end{table*}

\end{document}